\documentstyle[12pt]{article}\newcommand{\nc}{\newcommand}
\nc{\be}{\beta}
\nc{\th}{\theta} \nc{\dl}{\delta} \nc{\ph}{\varphi}
\nc{\ep}{\varepsilon}
\nc{\ov}{\over} \nc{\ra}{\rightarrow} \nc{\hf}{{1\ov2}}
\nc{\bq}{\begin{equation}} \nc{\eq}{\end{equation}}
\nc{\bZ}{{\bf Z}} \nc{\ba}{\left(\begin{array}{cc}}
\nc{\ea}{\end{array}\right)} \nc{\bR}{{\bf R}} \nc{\Dl}{\Delta}
\nc{\bT}{{\bf T}} \nc{\bC}{{\bf C}} \nc{\la}{\lambda}
\renewcommand{\sp}{\vspace{1ex}}
\nc{\noi}{\noindent}\nc{\hff}{\scriptstyle{\hf}}
\nc{\si}{\sigma} \nc{\pl}{\partial} \nc{\iy}{\infty} \nc{\ps}{\psi}
\nc{\et}{\eta}
\nc{\ch}{\raisebox{.4ex}{$\chi$}} \nc{\cd}{\cdots} \nc{\tn}{\otimes}
\begin{document} \mbox{}\vspace{-2ex}
\begin{center}{\large\bf Correlation Functions, Cluster Functions and
Spacing\\ \sp
Distributions for Random Matrices}\end{center}
\sp\begin{center}{{\bf Craig A. Tracy}\\
{\it Department of Mathematics and Institute of Theoretical Dynamics\\
University of California, Davis, CA 95616, USA\\
e-mail address: tracy@itd.ucdavis.edu}}\end{center}
\begin{center}{{\bf Harold Widom}\\
{\it Department of Mathematics\\
University of California, Santa Cruz, CA 95064, USA\\
e-mail address: widom@math.ucsc.edu}}\end{center}\sp

\begin{abstract} The usual formulas for the correlation functions in
orthogonal
and symplectic matrix models express them as quaternion determinants.
{}From this
representation one can deduce formulas for spacing probabilities in
terms of Fredholm
determinants of matrix-valued kernels. The derivations of the various
formulas are
somewhat involved. In this article we present a direct approach which
leads
immediately to scalar kernels for the unitary ensembles and matrix
kernels
for the orthogonal and symplectic ensembles,
and the representations of the correlation functions, cluster functions
and
spacing distributions in terms of them. \end{abstract}\sp

\renewcommand{\theequation}{1.\arabic{equation}}
\begin{center}{\bf 1. Introduction}\end{center}\sp

In the most common models of random matrices the eigenvalue distribution
is given by a
probability density
$P_N(x_1,\cd,x_N)$. If $F$ is a symmetric function of $N$
variables and the eigenvalues
are $\la_1,\cd,\la_N$ then the expected value of $F(\la_1,\cd,\la_N)$ is
given
by the formula
\[\int\cd\int\,F(x_1,\cd,x_N)\,P_N(x_1,\cd,x_N)\,dx_1\cd dx_N.\]
Here $dx$ denotes Lebesgue measure on a set which necessarily contains
the
eigenvalues (\bR\ if the matrices are Hermitian, \bT\ if the matrices
are
unitary, \bC\ if the matrices have general complex entries). The
function $P_N(x_1,\cd,x_N)$ gives the probability density that the
eigenvalues lie in
infinitesimal
neighborhoods of $x_1,\cd,x_N$. The $n$-point correlation function
$R_n(x_1,\cd,x_n)$ is defined by
\[R_n(x_1,\cd,x_n)={N!\ov(N-n)!}\int\cd\int\,P_N(x_1,\cd,x_N)\,dx_{n+1}\cd
dx_N.\]
It is, loosely speaking, the probability density that $n$ of
the eigenvalues, irrespective of order, lie in infinitesimal
neighborhoods of $x_1,\cd,x_n$.
(It is not a probability density in the strict sense
since its total integral equals $N!/(N-n)!$ rather than 1.)

Sometimes easier to compute are the $n$-point cluster functions
$T_n(x_1,\cd,x_n)$.
These are defined in terms of the $R_m(x_1,\cd,x_m)$ with $m\leq n$.
Reciprocally,
the $R_n(x_1,\cd,x_n)$ may be recovered from the $T_m(x_1,\cd,x_m)$ with
$m\leq n$. These will be discussed in Section 2.

Fundamental to the study of spacings between consecutive eigenvalues is
the quantity
$E(0;J)$ (the $E$ here does not represent expected value), which is the
probability
that the set $J$ contain no eigenvalues. More generally
$E(n_1,\cd,n_m;J_1,\cd,J_m)$
denotes the probability that each $J_i$ contains precisely $n_i$
eigenvalues. A related quantity is $P_J(x_1,\cd,x_n)$, the probability
density that the eigenvalues contained in $J$ lie in
infinitesimal intervals about $x_1,\cd,x_n$. These
are useful for the study of spacings between several eigenvalues.

All the functions $R_n$ and $T_n$, as well as the just-mentioned
probabilities, are
obviously expressible in terms of the density function $P_N$. But since
we generally
think of $N$ as large (indeed we are often interested in ``scaling
limits'' as
$N\ra\iy$) it is desirable to find alternative expressions for these
quantities which
do not increase in complexity with large $N$, expressions in which $N$
appears simply
as a parameter.

These have been found for some particularly important matrix
ensembles. They are ensembles of Hermitian matrices where $P_N$ has the
form
\[P_N(x_1,\cd,x_N)=c_N\prod_{j<k}|x_j-x_k|^{\be}\,\prod_jw(x_j),\]
where $\be$ equals 1, 2, or 4, $w(x)$ is a weight function and $c_N$ is
the normalization
constant required to make the total integral of $P_N$ equal to one; and
ensembles of
unitary matrices where $P_N$ has the form
\[P_N(x_1,\cd,x_N)=c_N\prod_{j<k}|e^{ix_j}-e^{ix_k}|^{\be},\]
and the integrals are taken over any interval of length $2\pi$. The
latter are Dyson's
circular ensembles. (Of course one could also
introduce a weight function here.) The terminology here can be
confusing.
The $\be=1, 2$ and 4 ensembles
are called orthogonal, unitary and symplectic ensembles, respectively,
because the underlying measures are invariant under actions of these
groups. In the
Hermitian case the $\be=1$ ensembles consist of real symmetric
matrices and the
$\be=4$ ensembles consist of self-dual Hermitian matrices. For a
discussion of these matters and the analogues for the circular
ensembles see \cite{M2}.

Here is a rough outline of how the
desired expressions were obtained when $\be=2$, which are the simplest
cases. For
Hermitian ensembles
\[P_N(x_1,\cd,x_N)=c_N\prod_{j<k}(x_j-x_k)^2\,\prod_jw(x_j).\]
On the right side we see the square of a Vandermonde determinant (for
the circular ensemble it is a product of Vandermonde determinants),
which is equal to
the determinant of the Vandermonde matrix times its transpose.
Performing
row and column operations and writing out the product, one obtains
a representation
\bq P_N(x_1,\cd,x_N)={1\ov N!}\,\det\,(K_N(x_j,x_k))_{j,k=1,\cd,N}\label{Prep}\eq
with certain kernels $K_N(x,y)$ which are expressed in terms of the
polynomials
orthonormal with respect to the weight function $w(x)$. These kernels
are shown to
satisfy a basic family of integral identities which are used to deduce
that
\bq R_n(x_1,\cd,x_n)=\det\,(K_N(x_j,x_k))_{j,k=1,\cd,n}.\label{Rform}\eq
By manipulating the $N$-fold integral that gives $E(0;J)$ and using
these expressions
for the $R_n$ one deduces that $E(0;J)$ equals the Fredholm determinant
of the kernel
$K_N(x,y)$ acting on $J$. One obtains expressions for the more general
quantities
$E(n_1,\cd,n_m;J_1,\cd,J_m)$ in terms of the same kernel. Observe that
$N$
appears only as a parameter in the kernel $K_N(x,y)$.

For $\be$ equal to 1 or 4 the situation is less simple. From the point
of view of
the above outline, the problem is to concoct a kernel satisfying
(\ref{Prep}) and the
family of integral identities which led to formula (\ref{Rform}).
This problem was solved by Dyson \cite{D} in the case of the circular
ensembles
and required the introduction of a new concept---the quaternion
determinant.
A quaternion-valued kernel was produced for which (\ref{Prep}) and
(\ref{Rform}) held if
all determinants were interpreted as quaternion determinants.
Subsequently it was
shown by Mehta \cite{M1} how to obtain analogous quaternion determinant
representations for the Gaussian  ensembles, the ensembles of
Hermitian matrices with weight function $w(x)=e^{-x^2}$,
using skew-orthogonal polynomials. This was generalized to general
weights by
Mehta and Mahoux \cite{MM}. (See also \cite{M2,F,NW}.) After
these quaternion determinant representations for the $R_n$
it is possible (by an argument to be found in A.7 of \cite{M2}, although
not explicitly
stated there) to deduce representations for $E(0;J)^2$ as Fredholm
determinants of
$2\times2$ matrix kernels \cite{TW}. These matrix kernels are the matrix
representations of
the quaternion kernels.

If all one is interested in is the spacing distributions then this is
certainly
a very roundabout way of obtaining Fredholm determinant
representations. The purpose of this
article is to present a direct approach, one which leads
immediately to the scalar kernels when $\be=2$ and the matrix kernels
when $\be=1$ and 4,
and the representations of the correlation functions, cluster functions
and
spacing distributions in terms of them. It uses neither quaternion
determinants nor a family of integral identities
for the kernels. (That the correlation functions are equal to quaternion
determinants
for the $\be=1$ and 4 ensembles becomes a consequence of the
representations.)

What we do use are the following three identities which represent
certain $N$-fold integrals
with determinant entries in terms of $N\times N$ or $2N\times 2N$
determinants with integral
entries:

\[\int\cd\int\,\det(\phi_j(x_k))_{j,k=1,\cd,N}\,\cdot
\det(\psi_j(x_k))_{j,k=1,\cd,N}\,
dx_1\cd dx_N\]
\bq=N!\,\det\Big(\int
\phi_j(x)\,\psi_k(x)\,dx\Big)_{j,k=1,\cd,N}.\label{2det}\eq\sp

\[\stackrel{\displaystyle{\int\cd\int}}{\scriptstyle{x_1\leq\cd\leq
x_N}}
\raisebox{2.5ex}{$\det(\phi_j(x_k))_{j,k=1,\cd,N}\,dx_1\cd dx_N$}\]
\bq={\rm Pf}\Big(\int\int{\rm
sgn}(y-x)\,\phi_j(x)\,\phi_k(y)\,dy\,dx\Big)_{j,k=1,\cd,N}.
\label{1det}\eq\sp

\[\int\cd\int\,\det(\phi_j(x_k)\ \ \psi_j(x_k))_{j=1,\cd,2N,\
k=1,\cd,N}\,
dx_1\cd dx_N\]
\bq=(2N)!\;{\rm
Pf}\Big(\int(\phi_j(x)\,\psi_k(x)-\phi_k(x)\,\psi_j(x))\,dx\Big)_
{j,k=1,\cd,2N}.
\label{4det}\eq
Here ``Pf'' denotes Pfaffian (its square
is the determinant) and (\ref{1det}) holds for even $N$ and must be
modified for odd
$N$. These hold for general measure spaces; in (\ref{1det}) the
space must be ordered. Identities (\ref{1det}) and (\ref{4det}) are due
to de Bruijn
\cite{deB}, who traces (\ref{2det}) as far back as 1883 \cite{A}.

In the application of (\ref{2det}) to the $\be=2$ case we are
led to a matrix whose $j,k$ entry equals $\dl_{j,k}$ plus the integral
of a product of a function of $j$ and a function of $k$. In our
application of
(\ref{1det}) and (\ref{4det}) to $\be=1$ and 4 the integrand is a sum of
two such products.
This, in a nutshell, is why $2\times2$ matrix kernels arise.
What we do in each case is use one of the identities to express
\bq\int\cd\int P_N(x_1,\cd,x_N)\,\prod_j(1+f(x_j))\,dx_1\cd
dx_N\label{Pfint}\eq
as a determinant or Pfaffian whose entries are given by one-dimensional
integrals. Manipulating the integrals leads to a
scalar or matrix kernel $K_N(x,y)$ such that the above $N$-fold integral
($\be=2$) or its
square
($\be=1$ or 4) is equal to $\det\,(I+K_N\,f)$. Here $K_N$ denotes the
operator with kernel
$K_N(x,y)$ and $f$ denotes multiplication by that function. Taking
$f=-\ch_J$ gives
immediately the representation of $E(0;J)$ or its square as
$\det\,(I-K_N\,\ch_J)$.
We also obtain representations for the more general quantities
$E(n_1,\cd,n_m;J_1,\cd,J_m)$.
Taking $f$ to be a linear combination of  delta functions leads to
representations
of the correlation and cluster functions in terms of the matrices
$\big(K(x_i,x_j)\big)$.
These are $2\times2$ block matrices if $\be=1$ or 4.

All the derivations follow the same pattern. First we write the
integrand in (\ref{Pfint})
in terms of a determinant or product of determinants. These
representation are
exactly as can be found in \cite{D} or \cite{M2}, for example. Then we
apply whichever of
(\ref{2det})--(\ref{4det}) is appropriate, and obtain a determinant
whose entries are
integrals. Finally, after writing the integrand as a matrix product if
necessary, we use
a general identity to express the last determinant as an operator
determinant.

The article is organized as follows. In section 2 we define the cluster
functions and
discuss their relationship with the correlation functions. In sections 3
and 4 we show
how the operator determinant representations of the
integrals (\ref{Pfint}) lead to formulas for the correlation and cluster
functions as
well as the spacing probabilities. In the following sections we derive
these
determinant representations for the $\be=2$ ensembles, the $\be=4$ and
$1$ circular
ensembles, and then the $\be=4$ and $1$ Hermitian ensembles. We do
things in this
order since the circular ensembles are simpler than the Hermitian
(because the
exponentials are eigenfunctions of the differentiation operator) and
$\be=4$ is simpler than $\be=1$ (because there is a single integral in
(\ref{4det}) and
a double integral in (\ref{1det})). In the final section we derive the
matrix kernels
for the $\be=4$ and $1$ Gaussian ensembles, which was the starting point
of \cite{TW}.

We emphasize that there are almost no new results in this
article---perhaps the determinant formula (\ref{spacings})
for the spacing probabilities and the more general
form for the
matrix kernels obtained in secs.~8 and 9. It is the methods
used to derive them which are new
and, we believe, show how they form a coherent whole.\sp

\setcounter{equation}{0}\renewcommand{\theequation}{2.\arabic{equation}}
\begin{center}{\bf 2. Correlation and Cluster Functions}\end{center}\sp

The $n$-point cluster functions $T_n(x_1,\cd,x_n)$ are
defined as follows. For each non-empty subset $S$ of $\{1,\cd,N\}$ write
\[R_S=R_n(x_{i_1},\cd,x_{i_n})\ \ \ {\rm if}\ S=\{i_1,\cd,i_n\}.\]
Then
\bq T_n(x_1,\cd,x_n)=\sum\,(-1)^{n-m}\,(m-1)!\,R_{S_1}\cd
R_{S_m},\label{T}\eq
where the sum runs over all $m\geq1$ and, for each $m$, over all
partitions of
$\{1,\cd,n\}$ into nonempty
subsets $S_1,\cd,S_m$. If we know the $T_n$ then the $R_n$ may be
recovered by use
of the reciprocal formula
\bq R_n(x_1,\cd,x_n)=\sum\,(-1)^{n-m}\,T_{S_1}\cd T_{S_m}.\label{R}\eq
This is most easily seen by looking at the relation between generating
functions for
these quantities.

Let $A$ be a formal power series in variables
$z_1,\cdots,z_N$ without constant
term. For each nonempty subset $S$ of $\{1,\cdots,N\}$ let $A_S$ denote
the
coefficient of $\prod_{i\in S}z_i$ in $A$. Let
\[f(z)=\sum_{m=1}^{\infty}f_m\,z^m\]
be a formal power series in a single variable $z$ without constant term,
and
set $B=f(A)$. For each $S$ we have
\bq B_S=\sum m!\,f_m\,A_{S_1}\cd A_{S_m},\label{id1}\eq
where the sum is taken over all $m\geq1$ and, for each $m$, all
partitions of $S$
into nonempty subsets ${S_1,\cd,S_m}$.
If $f_1\neq0$ then $f(z)$ has a formal inverse $g(z)$. Since $A=g(B)$ it
follows from the above that for each $S$
\bq A_S=\sum m!\,g_m\,B_{S_1}\cd B_{S_m}.\label{id2}\eq

We apply these fact to $f(z)=-\log(1+z),\ g(z)=e^{-z}-1$, taking any
formal power series
$A$ such that $A_S=R_n(x_{i_1},\cd,x_{i_n})$ when $S=\{i_1,\cd,i_n\}$.
Since
$f_m=(-1)^m/m$ in this case, we see by
definition (\ref{T}) that with the same $S$ we have $B_S=
(-1)^n\,T_n(x_{i_1},\cd,x_{i_n})$. Since $g_m=(-1)^m/m!$ (\ref{id2})
gives the
reciprocal formula (\ref{R}).

Observe that $R_n(y_1,\cd,y_n)$ equals
the coefficient of $z_1\cd z_n$ in the
expansion about $z_1=\cd=z_n=0$ of
\bq\int\cdots\int\ P_N(x_1,\cdots,x_N)\ \prod_j[1+\sum_{r=1}^n
z_r\,\dl(x_j-y_r)]\ dx_1\cdots dx_N.\label{Rn}\eq
Here $(1+\sum_{r=1}^nz_r\,\dl(x-y_r))\,dx$ denotes Lebesgue measure plus
masses $z_r$ at the points $y_r$. Thus we may take $A(z_1,\cd,z_n)$ to
be this
integral minus 1. Since $B=-\log(1+A)$ we see, after recalling the
relationship
$T_S=(-1)^{|S|}B_S$, that $T_n(y_1,\cd,y_n)$ equals
$(-1)^{n+1}$ times the coefficient of $z_1\cd z_n$ in
the expansion about $z_1=\cd=z_n=0$ of
\bq\log\Big(\int\cdots\int\ P_N(x_1,\cdots,x_N)
\ \prod_j[1+\sum_{r=1}^nz_r\,\dl(x_j-y_r)]\ dx_1\cdots
dx_N\Big).\label{Tn}\eq\sp

\setcounter{equation}{0}\renewcommand{\theequation}{3.\arabic{equation}}
\begin{center}{\bf 3. Formulas for the correlation and cluster
functions}\end{center}\sp

Suppose we have proved that for each $\be=2$ ensemble there is a kernel
$K_N(x,y)$
such that for general $f$
\bq\int\cd\int P_N(x_1,\cd,x_N)\,\prod_j(1+f(x_j))\,dx_1\cd dx_N
=\det\,(I+K_N\,f).\label{2Pf}\eq
As before $K_N$ denotes the operator on $L^2$ with kernel $K_N(x,y)$
and $f$ as an operator denotes multiplication by that function.

Take $f(x)=\sum_{r=1}^n z_r\,\dl(x-y_r)$ first. Then the operator
$I+K_N\,f$ becomes the
matrix
\bq\Big(\dl_{r,s}+K_N(y_r,y_s)\,z_s\Big)_{r,s=1,\cd,n}.\label{Kmatrix}\eq
This may be seen by passing to a limit or observing (as we shall) that
this is what
happens during the derivation. It is easy to see that the coefficient of
$z_1\cd z_n$ in the
expansion of the determinant of (\ref{Kmatrix}) equals
$\det\Big(K_N(y_r,y_s)\Big)$, and so
by (\ref{Rn}) we obtain formula (\ref{Rform}) for the correlation
functions.

For the cluster functions we use (\ref{Tn}) and the general expansion
\[\log\det\,(I+K)={\rm tr}\,\log\,(I+K)=\sum_{m=1}^{\iy}{(-1)^{m+1}\ov
m}\ {\rm tr}\ K^m\]
to deduce
\[T_n(y_1,\cd,y_n)={1\ov n}\,\sum_{\si}K_N(y_{\si1},y_{\si2})\cd
K_N(y_{\si n},y_{\si1}),\]
the sum taken over all permutations $\si$ of $\{1,\cd,n\}$.

Turning now to the $\be=1$ and 4 ensembles, suppose we can show that for
each of them
there is a matrix kernel $K_N(x,y)$ such that
\bq\int\cd\int P_N(x_1,\cd,x_N)\,\prod_j(1+f(x_j))\,dx_1\cd dx_N
=\sqrt{\det\,(I+K_N\,f)}\,.\label{1Pf}\eq
Then taking $f(x)=\sum_{r=1}^N z_r\,\dl(x-y_r)$ we see by (\ref{Rn})
that
$R_n(y_1,\cd,y_n)$
equals the coefficient of $z_1\cd z_n$ in the expansion of
\bq\sqrt{\det\,\Big(\dl_{r,s}+
K_N(y_r,y_s)\,z_s\Big)_{r,s=1,\cd,n}}\,.\label{1R}\eq
This coefficient is not nearly as simple as in the $\be=2$ case. However
(\ref{Tn}) shows that
$T_n(y_1,\cd,y_n)$ equals
$(-1)^{n+1}$ times the coefficient of $z_1\cd z_n$ in the expansion of
\[\hf\,\log\det\Big(\dl_{r,s}+K_N(y_r,y_s)\,z_r\Big).\]
Thus we obtain
\[T_n(y_1,\cd,y_n)
={1\ov 2n}\,\sum_{\si}{\rm tr}\ K_N(y_{\si1},y_{\si2})\cd K_N(y_{\si
n},y_{\si1}),\]
which is hardly more complicated than for $\be=2$.
To obtain formulas for the correlation functions we
use (\ref{R}) together with our expressions for the $T_n$. Of course the
same formulas
must result if we take the appropriate coefficients in (\ref{1R}).

It turns out that the $2\times2$ block matrices $\Big(K_N(y_r,y_s)\Big)$
are always
self-dual. This means that the
matrices on the diagonal are transposes of each other and
the matrices off the diagonal are antisymmetric. Because of this the
expression for
the correlation function given by (\ref{R}) shows that (\ref{Rform})
holds when the
determinant is interpreted as a quaternion determinant. For a discussion
of
this point see \cite{M2}, sec.~6.2.\sp

\setcounter{equation}{0}\renewcommand{\theequation}{4.\arabic{equation}}
\begin{center}{\bf 4. Formulas for the spacing
probabilities}\end{center}\sp

We obtain the formulas first assuming  that $\be=2$ and (\ref{2Pf})
holds.
To evaluate $E(0;J)$ we must integrate
$P_N(x_1,\cd,x_N)$ over the region where all the $x_i$ lie in the
complement of $J$.
Thus we simply take $f=-\ch_J$ and obtain
\[ E(0;J)=\det\,(I-K_N\,\ch_J).\]
More generally, to
evaluate $E(n_1,\cd,n_m;J_1,\cd,J_m)$ we integrate
$P_N(x_1,\cd,x_N)$ over the region where $n_i$ of the $x_j$ lie in $J_i$
and all the other
$x_j$ lie outside the union of the $J_i$. We see that
$E(n_1,\cd,n_m;J_1,\cd,J_m)$
equals the coefficient of $(\la_1+1)^{n_1}\cd(\la_m+1)^{n_m}$
in the expansion about $\la_1=\cd=\la_m=-1$ of
\[\int\cdots\int\ P_N(x_1,\cdots,x_N)\ \prod_j[1+\sum_{i=1}^m
\la_i\,\ch_{J_i}(x_j)]\ dx_1\cdots dx_N.\]
Therefore we take $f(x)=\sum_{i=1}^m\la_i\,\ch_{J_i}(x)$ in (\ref{2Pf})
and deduce that
\[ E(n_1,\cd,n_m;J_1,\cd,J_m)\]
\bq={1\ov n_1!\cd n_m!}\,{\pl^{\sum
n_i}\ov\pl\la_1^{n_1}\cdots\pl\la_m^{n_m}}
\det\Big(I+K_N\,\sum\la_i\,
\ch_{J_i}\Big)\Big|_{\la_1=\cd=\la_m=-1}.\label{En}\eq

To see the connection with spacings between eigenvalues, consider the
quantity
\[[E(0;\,(x+\Dl x,y))-E(0;\,(x,y))]-[E(0;\,(x+\Dl x,y+\Dl
y))-E(0;\,(x,y+\Dl y))],\]
where $\Dl x$ and $\Dl y$ are positive.
The difference on the left equals the probability that there is no
eigenvalue in
$(x+\Dl x,y)$ but there is an eigenvalue in the larger interval $(x,y)$,
which is the
same as saying that there is no eigenvalue in $(x+\Dl x,y)$ but there is
an eigenvalue in
$(x,x+\Dl x)$.
Similarly the difference on the right equals the probability that there
is no eigenvalue in
$(x+\Dl x,y+\Dl y)$ but there is an eigenvalue in $(x,x+\Dl x)$.
Therefore the
difference equals the probability that there is no eigenvalue in $(x+\Dl
x,y)$ but
there is an eigenvalue in both $(x,x+\Dl x)$ and $(y,y+\Dl x)$.
Hence if we divide this difference by $\Dl x\,\Dl y$ and take the limit
as
$\Dl x,\;\Dl y\ra0$ we obtain the joint probability density for
consecutive eigenvalues
to lie in infinitesimal intervals about $x$ and $y$. On the other hand
this
same limit equals
\[-{d^2\ov dy\,dx}E(0;\,(x,y)).\]
Therefore this is the formula for the joint probability density, and
explains why the spacing
distributions are intimately connected with Fredholm determinants.

To obtain the conditional probability density that, given an eigenvalue
at $x$,
the next eigenvalue lies in an infinitesimal neighborhood of $y$, we
must divide the above by the probability density that there is an
eigenvalue in an
infinitesimal neighborhood of $x$. This is the 1-point correlation
function $R_1(x,x)$
which, we know, equals $K_N(x,x)$. Therefore this conditional
probability density
equals
\[-{1\ov K_N(x,x)}\,{d^2\ov dy\,dx}E(0;\,(x,y)).\]

We now compute $P_J(x_1,\cd,x_n)$, the joint probability density that
the
eigenvalues contained in an interval $J$ lie in
infinitesimal intervals about $x_1,\cd,x_n$.
We observe first that the probability that a small interval contains
more than one
eigenvalue goes to zero as the square of the length of the interval.
So we shall apply (\ref{En}) to small intervals about the $x_i$ (taking
these $n_i=1$) and the intervals constituting the rest of $J$ (taking
these $n_i=0$),
divide by the product of
the lengths of the small intervals and let their lengths tend to 0.
We denote the small intervals by $J_1,\cd,J_n$ and the rest of $J$ by
$J'$. Denote by
$E_{J_1,\cd,J_n}$ the probability that each $J_i$ contains one
eigenvalue and $J'$ none.
By (\ref{En}) we have
\[E_{J_1,\cd,J_n}={\pl^n\ov\pl\la_1\cdots\pl\la_n}
\det\Big(I-K_N\,\ch_{J'}+K_N\,\sum_{i=1}^n\la_i\,\ch_{J_i}\Big)\Big|
_{\la_1=\cd=\la_n=-1}\]
\[={\pl^n\ov\pl z_1\cdots\pl z_n}
\det\Big(I-K_N\,\ch_{J}+K_N\,\sum_{i=1}^n
z_i\,\ch_{J_i}\Big)\Big|_{z_1=\cd=z_n=0}.\]
We factor the operator in this expression as
\[(I-K_N\,\ch_{J})\,(I+R_J\,\sum_{i=1}^n z_i\,\ch_{J_i}),\]
where $R_J$ is the resolvent kernel of $K_N\,\ch_{J}$, the kernel of
$(I-K_N\,\ch_{J})^{-1}\,K_N\,\ch_J$. Determinants multiply, and the
determinant
of the first factor equals $E(0;J)$, as we know. Thus
\[E_{J_1,\cd,J_n}=E(0;J)\,{\pl^n\ov\pl z_1\cdots\pl z_n}
\det\Big(I+R_J\,\sum_{i=1}^n z_i\,\ch_{J_i}\Big)\Big|_{z_1=\cd=z_n=0}.\]

Now we can proceed as for the correlation and cluster functions. If
\[A(z_1,\cd,z_n)=\det\Big(I+R_J\,\sum_{i=1}^n z_i\,\ch_{J_i}\Big)-1,\]
\[B(z_1,\cd,z_n)=\log\,\det\Big(I+R_J\,\sum_{i=1}^n
z_i\,\ch_{J_i}\Big),\]
then $E_{J_1,\cd,J_n}/E(0;J)$ equals the coefficient of $z_1\cd z_n$ in
the expansion
of $A$ about $z_1=\cd=z_n=0$ and $A=e^B-1$. The coefficients of $A$ are
determined from
the coefficients of $B$ by formula (\ref{id2}), which gives in this case
\[A_S=\sum B_{S_1}\cd B_{S_m},\]
where the sum is taken over all partitions $\{S_1,\cd,S_m\}$ of $S$ into
nonempty subsets. To evaluate
\[B_S=\prod_{i\in S}{\pl\ov\pl z_i}
\log\,\det\,\Big(I+R_J\,\sum_{i\in S}
z_i\,\ch_{J_i}\Big)\Big|_{z_1=\cd=z_n=0}\]
we use the general facts for an operator function $K(z)$ that
\[{d\ov dz}\log\,\det\,K(z)={\rm tr}\, K(z)^{-1}\,K'(z),\ \ \ {d\ov
dz}K(z)^{-1}
=-K(z)^{-1}\,K'(z)\,K(z)^{-1}.\]
We apply this consecutively to the derivatives with respect to the $z_i$
and
find that if $S=\{i_1,\cd, i_k\}$ then
\[B_S={(-1)^{k-1}\ov k}\,\sum_{\si}{\rm tr}\ R_J\,\ch_{J_{i_{\si1}}}\cd
R_J\,
\ch_{J_{i_{\si k}}},\]
where $\si$ runs over all permutations of $\{1,\cd,k\}$.

Eventually we have to divide the various products $B_{S_1}\cd B_{S_m}$
by the product of the
lengths of all the $J_i$ and take the
limit as these lengths tend to 0. We may consider separately each $B_S$
and
its corresponding intervals since the $S_i$ are disjoint. So we compute
\[{\rm tr}\ R_J\,\ch_{J_{i_1}}\cd R_J\,\ch_{J_{i_k}}=
\int_{J_{i_k}}\cd\int_{J_{i_1}}R_J(x_{i_k},x_{i_1})\,R_J(x_{i_1},x_{i_2})\cd
R_J(x_{i_{k-1}},x_{i_k})\,dx_{i_1}\cd dx_{i_k}.\]
If $J_{i_1},\cd, J_{i_k}$ are small intervals about $x_{i_1},\cd,
x_{i_k}$ then dividing
by the product of their lengths and letting the lenths tend to 0 gives
in the limit
\[R_J(x_{i_k},x_{i_1})\,R_J(x_{i_1},x_{i_2})\cd
R_J(x_{i_{k-1}},x_{i_k}).\]

Now we would have obtained exactly the same results if, instead of
starting with the
determinant of the operator $I+R_J\,\sum z_i\,\ch_{J_i}$,
differentiating with respect to the $z_i$, setting all $z_i=0$,
dividing and taking a limit, we had simply started
with the determinant of the matrix
\[\Big(\dl_{i,j}+R_J(x_i,x_j)\,z_j\Big)_{i,j=1,\cd,n}\]
and differentiated with respect to the $z_i$ and set all $z_i=0$.
It follows that the the ratio $P_J(x_1,\cd,x_n)/E(0;J)$ equals
\[{\pl^n\ov\pl z_1\cdots\pl z_n}
\det\Big(\dl_{i,j}+R_J(x_i,x_j)\,z_j\Big)\Big|_{z_1=\cd=z_n=0}.\]
Therefore we have obtained the formula
\bq P_J(x_1,\cd,x_n)=E(0;J)\, \det\Big(R_J(x_i,x_j)\Big)_{i,j=1\cd
n}.\label{spacings}\eq
Note the similarity to formula (\ref{Rform}) for the correlation
functions: to obtain the
formula for the ratio $P_J(x_1,\cd,x_n)/E(0;J)$ we replace the kernel
$K_N(x,y)$ by the kernel $R_J(x,y)$.

We mention that rather more complicated
formulas for these probability densities were obtained by Mehta and des
Cloizeaux
\cite{MdC}. (See also \cite{M2}.) These involve the eigenvalues and
eigenfunctions
of the operator $K_N\,\ch_{J}$ but they must must be equivalent to
(\ref{spacings}).

For the spacing probabilities when $\be=1$ or 4 we must make
modifications as at
the end of the previous section. We have in this case
\[E(0;J)^2=\det\,(I-K_N\,\ch_J)\]
and the ratio $P_J(x_1,\cd,x_n)/E(0;J)$
is again given by the same formulas as for the correlation functions but
with $K_N(x,y)$
replaced by $R_J(x,y)$. One can show that the block matrices
$\big(R_J(x_i,x_j)\big)$ that
arise here are also self-dual, and this
implies that the formula for $P_J(x_1,\cd,x_n)$ is as above if the
determinant is interpreted as a quaternion determinant.\sp

\setcounter{equation}{0}\renewcommand{\theequation}{5.\arabic{equation}}
\begin{center}{\bf 5. The {\boldmath$\be=2$}\ matrix
ensembles}\end{center}\sp

Consider the Hermitian case first. We use (\ref{2det}) to express
\bq\int\cd\int\prod_{j<k}(x_j-x_k)^2\,\prod_jw(x_j)\,\prod_j(1+f(x_j))\,dx_1\cd
dx_N
\label{2fdet}\eq
as a determinant. Since
\[\prod_{j<k}(x_j-x_k)=\det\,(x_k^j)_{j=0,\cd,N-1,\ k=1,\cd,N}\]
applying (\ref{2det}) with $\phi_j(x)=\psi_j(x)=x^j$ and with the
measure $w(x)\,(1+f(x))\,dx$
shows that this equals $N!$ times
\[\det\Big(\int x^{j+k}\,w(x)\,(1+f(x))\,dx\Big)_{j,k=0,\cd,N-1}.\]
If we denote by $\{\ph_j(x)\}$ the sequence obtined by orthonormalizing
the sequence
$\{x^j\,w(x)^{\hf}\}$ then we see that the above is equal, except for a
different constant
factor depending only on $N$, to
\[\det\Big(\int\ph_j(x)\,\ph_k(x)\,(1+f(x))\,dx\Big)_{j,k=0,\cd,N-1}\]
\bq=\det\Big(\dl_{j,k}+\int\ph_j(x)\,\ph_k(x)\,f(x)\,dx\Big)_{j,k=0,\cd,N-1}.
\label{phjk}\eq

Now we use something which is needlessly fancy but which is very
useful, namely the general relation $\det\,(I+AB)=
\det\,(I+BA)$ for arbitrary Hilbert-Schmidt operators $A$ and $B$. They
may act between
different spaces as long as the products make sense. In our case we take
$A$ to be the
operator from $L^2$ to $\bC^N$ with kernel $A(j,x)=\ph_j(x)\,f(x)$ and
$B$
the operator from $\bC^N$ to $L^2$ with kernel $B(x,j)=\ph_j(x)$. Then
\[AB(j,k)=\int\ph_j(x)\,\ph_k(x)\,f(x)\,dx,\]
so (\ref{phjk}) equals $\det\,(I+AB)$. Now $BA$ is the operator on $L^2$
with kernel $K_N(x,y)\,f(y)$ where
\[K_N(x,y)=\sum_{k=0}^{N-1}\ph_k(x)\,\ph_k(y).\]
Therefore (\ref{phjk}) also equals
$\det(I+K_N\,f)$.

Hence we have shown for these ensembles
that (\ref{2Pf}) holds up to a constant factor depending only on $N$.
In fact the constant factor now must be 1, as is seen by setting $f=0$,
and so (\ref{2Pf})
is true as it stands.

It is easily seen directly that the nonzero eigenvalues of $K_Nf$
are the same as those of the matrix
$(\int\ph_j(x)\,\ph_k(x)\,f(x)\,dx)$, for any eigenfunction must be
a linear combination of the $\ph_j$ and solving for the coefficients
leads directly to the
matrix. So the fact $\det(I+AB)=\det(I+BA)$ is very simple here.

When the measure $f(x)\,dx$ is discrete, in other words when $f(x)=
\sum_{r=1}^nz_r\,\dl(x-y_r)$, the integral in the right side
of (\ref{phjk}) is replaced by
$\sum_{r=1}^nz_r\,\ph_j(y_r)\,\ph_k(y_r)$. If we set
\[A(j,r)=z_r\,\ph_j(y_r),\ \ B(r,j)=\ph_j(y_r),\ \ \ \ \ (j=0,\cd,N-1,\
\ r=1,\cd,n)\]
then our matrix is $AB$, and
\[BA(r,s)=K_N(y_r,y_s)\,z_s,\ \ \ \ (r,s=1,\cd,n)\]
with $K_N$ as before. Thus in (\ref{2Pf}) the operator $I+K_N\,f$ is
replaced by the
matrix (\ref{Kmatrix}), as claimed. (The same argument will hold for the
other
ensembles, and there will be no need to repeat it.)

For the circular $\be=2$ ensemble (\ref{2fdet}) is replaced by
\[\int\cd\int\,\prod_{j<k}(e^{-ix_j}-e^{-ix_k})\,\prod_{j<k}(e^{ix_j}-e^{ix_k})
\,\prod_j(1+f(x_j))\,dx_1\cd dx_N.\]
In the application of (\ref{2det}) we take $\phi_j(x)=
e^{-ix_j}/\sqrt{2\pi},\ \psi_j(x)=e^{ix_j}/\sqrt{2\pi}$ so that the
analogue of the matrix
in (\ref{phjk}) becomes
\[{1\ov2\pi}\int\,e^{i(k-j)x}\,f(x)\,dx.\]
(The $1/2\pi$ factor is needed to obtain $\dl_{j,k}$ when $f=0$.) Now we
set
\[A(j,x)={1\ov2\pi}e^{-i(j-{N-1\ov2})x}\,f(x),\ \ \
B(x,j)=e^{i(j-{N-1\ov2})x},\]
(the extra exponential factors yield a simpler formula), so that
the above matrix equals $AB$ while $BA$ equals the operator with  kernel
$K_N(x,y)\,f(y)$ where now
\[K_N(x,y)={1\ov2\pi}\sum_{k=0}^{N-1}e^{i(k-{N-1\ov2})(x-y)}=
{1\ov2\pi}{\sin\,{N\ov2}(x-y)\ov\sin\,{1\ov2}(x-y)}.\]
Thus (\ref{2Pf}) holds for the circular ensemble with this replacement
for $K_N$.\sp

\setcounter{equation}{0}\renewcommand{\theequation}{6.\arabic{equation}}
\begin{center}{\bf 6. The {\boldmath$\be=4$}\ circular
ensemble}\end{center}\sp

We begin with the evaluation of
\bq\int\cd\int\prod_{j<k}|e^{ix_j}-e^{ix_k}|^4\,\prod_j(1+f(x_j))\,dx_1\cd
dx_N
\label{4fdet}\eq
as a Pfaffian. It is a pretty fact that
\bq\prod_{j<k}(x_j-x_k)^4=\det\,\Big(x_k^j\ \ jx_k^{j-1}\Big)_
{j=0,\cd,2N-1,\ k=1,\cd,N}.\label{pretty}\eq
This is seen by writing the product representation of the Vandermonde
determinant
\[\det\,\Big(x_k^j\ \ y_k^{j}\Big)_{j=0,\cd,2N-1,\ k=1,\cd,N}\]
then differentiating with respect to each $y_k$ and setting $y_k=x_k$.

If we replace each $x_k$ by $e^{ix_k}$ and use the fact that
\bq|e^{ix_j}-e^{ix_k}|=i\,e^{-i{x_j+
x_k\ov2}}\,(e^{ix_j}-e^{ix_k})\label{abs}\eq
if $x_j\leq x_k$ (both lying in the same interval of length $2\pi$) we
see that
\[\prod_{j<k}|e^{ix_j}-e^{ix_k}|^4=e^{-2i(N-1)\sum
x_j}\,\prod_{j<k}(e^{ix_j}-e^{ix_k})^4.\]
By (\ref{pretty}) this is equal to
\[e^{-2i(N-1)\sum x_j}\,\det\,\Big(e^{ijx_k}\ \
je^{i(j-1)x_k}\Big)_{j=0,\cd,2N-1,\ k=1,\cd,N}
=\det\,(e^{i(j-N+\hf)x_k}\ \ je^{i(j-N+\hf)x_k}\Big)\]
\[=\det\,(e^{i(j-N+\hf)x_k}\ \
(j-N+{\scriptstyle{\hf}})e^{i(j-N+\hf)x_k}\Big)
=\det\,(e^{ipx_k}\ \ pe^{ipx_k}\Big),\]
where in the last determinant $k=1,\cd,N$ as before but $p$ runs through
the half integers
$-N+\hf,\,-N+{3\ov2},\cd,\,N-\hf$.

By the symmetry of the integrand in (\ref{4fdet}) the integral equals
$N!$ times the
integral over this region $x_1\leq\cd\leq x_N$.
So we can use formula (\ref{4det}) to deduce that the square of
(\ref{4fdet}) equals a constant depending only on $N$ times the
$2N\times2N$ determinant
\[\det\,\Big(\int(q-p)\,e^{i(p+q)x}\,(1+f(x))\,dx\Big).\]
Both indices $p$ and $q$ run over the half-integers $-N+\hf,\cd,N-\hf$.
If we reverse the order of the rows and divide each column by its index
$q$ we see
that this determinant is equal to another constant depending only on $N$
times
\[\det\,\Big({1\ov4\pi}\int(1+{p\ov q})\,e^{i(-p+q)x}\,(1+f(x))\,dx\Big)
=\det\,\Big(\dl_{p,q}+{1\ov4\pi}\int(1+{p\ov
q})\,e^{i(-p+q)x}\,f(x)\,dx\Big).\]

Here we see the sum of products $e^{-ipx}\,e^{iqx}+p\,e^{-ipx}\,{1\ov
q}\,e^{iqx}$
referred to in the introduction, and we write it as a matrix product
\[\left(\begin{array}{cc}e^{-ipx}&ipe^{-ipx}\end{array}\right)
\,\left(\begin{array}{c}e^{iqx}\\{1\ov iq}\,e^{iqx}\end{array}\right).\]
(The reason for the insertions of $i$ will become apparent.) Thus if we
set
\[A(p,x)={1\ov4\pi}\,f(x)\,
\left(\begin{array}{cc}e^{-ipx}&ip\,e^{-ipx}\end{array}\right),
\ \ B(x,q)=\left(\begin{array}{c}e^{iqx}\\{1\ov
iq}\,e^{iqx}\end{array}\right)\]
then the above matrix is $I+AB$. In this case $BA$ is the integral
operator with
matrix kernel $K_N(x,y)\,f(y)$ where
\[K_N(x,y)=\left(\begin{array}{cc}{1\ov4\pi}\sum
e^{ip(x-y)}&{1\ov4\pi}\sum ip\,e^{ip(x-y)}\\\\
{1\ov4\pi}\sum {1\ov ip}\,e^{ip(x-y)}&{1\ov4\pi}\sum
e^{ip(x-y)}\end{array}\right).\]
If we write
\[S_N(x)={1\ov2\pi}\sum_p\,e^{ipx}={1\ov2\pi}{\sin Nx\ov\sin\hf x},\ \ \
DS_N(x)=
{d\ov dx}S_N(x),\ \ \ IS_N(x)=\int_0^xS_N(y)\,dy,\]
then
\bq K_N(x,y)=\hf\left(\begin{array}{cc}S_N(x-y)&DS_N(x-y)\\\\
IS_N(x-y)&S_N(x-y)\end{array}\right).\label{4Kcirc}\eq\sp
Note the relationship between $S_N(x-y)$ and the kernel $K_N(x,y)$ which
arises in the
$\be=2$ ensemble: the former equals the latter with $N$ replaced by
$2N$.

Recall that we have been computing, not the integral (\ref{4fdet}), but
its
square. Thus we have shown for this ensemble and with this $K_N$ that
(\ref{1Pf})
holds up to a constant factor depending only
on $N$. As before, taking $f=0$ shows that the constant factor equals
1.\sp

\setcounter{equation}{0}\renewcommand{\theequation}{7.\arabic{equation}}
\begin{center}{\bf 7. The {\boldmath$\be=1$}\ circular
ensemble}\end{center}\sp

We assume here that $N$ is even and begin with the evaluation of
\bq\int\cd\int\prod_{j<k}|e^{ix_j}-e^{ix_k}|\,\prod_j(1+f(x_j))\,dx_1\cd
dx_N.
\label{1fdet}\eq
Using (\ref{abs}) we see that in the region $x_1\leq\cd\leq x_N$,
\[\prod_{j<k}|e^{ix_j}-e^{ix_k}|=i^{-N(N-1)}\,e^{-i{N-1\ov2}\sum
x_j}\,\prod_{j<k}(e^{ix_k}-e^{ix_j})\]
\[=i^{-N(N-1)}\,e^{-i{N-1\ov2}\sum
x_j}\,\det\Big(e^{ijx_k}\Big)_{j=0,\cd,N-1,\ k=1,\cd,N}
=i^{-N(N-1)}\,\det\Big(e^{ipx_k}\Big),\]
where in the last determinant $k=1,\cd,N$ and $p$ runs through the half
integers from
$-{N\ov2}+\hf$ to ${N\ov2}-\hf$, as before but with $N$ replaced by
$N/2$.
The $i$-factor in the last term equals $\pm1$.
We now apply (\ref{1det}) with the measure $(1+f(x))\,dx$, using the
fact that
$\prod|e^{ix_j}-e^{ix_k}|$ is a symmetric function, and conclude that
the square of
(\ref{1fdet}) is a constant depending only on $N$ times
\[\det\Big(\int\int{\rm
sgn}(y-x)\,e^{ipx}\,e^{iqy}\,(1+f(x))\,(1+f(y))\,dy\,dx\Big).\]
First, we replace $p$ by $-p$ as before.
Then we set $\ep(x)=\hf\,{\rm sgn}(x)$, and replace ${\rm sgn}(y-x)$ by
$\ep(x-y)$ in the
determinant. These things just change the multiplying constant. We shall
also use the notation
$e_p(x)$ for the normalized exponential $e^{ipx}/\sqrt{2\pi}$. With
these modifications
the above becomes
\[\det\Big(\int\int\ep(x-y)\,e_{-p}(x)\,e_q(y)\,
(1+f(x))\,(1+f(y))\,dy\,dx\Big).\]

Denote by $\ep$ the operator with kernel $\ep(x-y)$. Observe that $\ep$
is antisymmetric
and $\ep e_p=e_p/ip$.
(The latter uses the fact that $p$ is an integer plus $\hf$.) Writing
\[(1+f(x))\,(1+f(y))=1+f(x)+f(y)+f(x)\,f(y)\]
and using the fact that $\ep$ is antisymmetric, we see that the double
integral equals
\[{1\ov iq}\dl_{p,q}+\int\,[{1\ov iq}\,f(x)\,e_{-p}(x)\,e_q(x)\,dx+
{1\ov ip}\,f(x)\,e_{-p}(x)\,e_q(x)-f(x)\,e_q(x)\,\ep
(fe_{-p})(x)]\,dx,\]
so the determinant equals another constant depending on $N$ times that
of the matrix
with $p,q$ entry
\[\dl_{p,q}+\int\,[f(x)\,e_{-p}(x)\,e_q(x)+
{q\ov p}\,f(x)\,e_{-p}(x)\,e_q(x)-iq\,f(x)\,e_q(x)\,\ep
(fe_{-p})(x)]\,dx.\]
At this point setting $f=0$ shows that the multiplying constant equals
1.

We write the integrand in the $p,q$ entry of the matrix as a matrix
product
\[\left(\begin{array}{ccc}{1\ov ip}\,f(x)\,e_{-p}(x)-f(x)\,\ep
(fe_{-p})(x)&&
f(x)\,e_{-p}(x)\end{array}\right)
\ \left(\begin{array}{c}iq\,e_q(x)\\\\e_q(x)\end{array}\right)\]
and apply $I+AB\ra I+BA$ as before to replace this by a matrix kernel.
For
simplicity of notation we shall write out the operator rather than the
kernel and
use the notation $a\tn b$ for the operator with kernel $a(x)\,b(y)$. The
operator $I+BA$
is equal to
\[\ba I+\sum\,e_p\tn fe_{-p}-\sum\,ip\,e_p\tn
f\,\ep(fe_{-p})&\sum\,ip\,e_p\tn fe_{-p}\\\\
\sum\,{1\ov ip}\,e_p\tn fe_{-p}-\sum\,e_p\tn
f\,\ep(fe_{-p})&I+\sum\,e_p\tn fe_{-p}.\ea\]

Now it is an easy fact that for an operator $A$ we have $(a\tn b)\,A=
a\tn (A^t\,b)$. Therefore the second sums in the entries of the first
column of the
matrix, with their minus signs, equal the corresponding entries of the
second column
right-multiplied by the operator $\ep\,f$. (Here $f$ denotes
multiplication by the function,
which is symmetric, and we used the fact that $\ep$ is antisymmetric.)
It follows that the
matrix may be written as the product
\[\ba I+\sum\,e_p\tn fe_{-p}&\sum\,ip\,e_p\tn fe_{-p}\\\\
\sum\,{1\ov ip}\,e_p\tn fe_{-p}-\ep f&I+\sum\,e_p\tn fe_{-p}\ea
\ \ba I&0\\\\\ep\,f&I\ea.\]
The operator on the right has determinant 1 and so the determinant of
the product
equals the determinant of the operator on the left, i.e., the
determinant of
\[I+\ba \sum\,e_p\tn e_{-p}&\sum\,ip\,e_p\tn e_{-p}\\\\
\sum\,{1\ov ip}\,e_p\tn e_{-p}-\ep&\sum\,e_p\tn e_{-p}\ea\,f.\]
If we recall the definition of $e_p$ and the notation of the last
section we see that
(\ref{1Pf}) holds for this ensemble when $K_N$ has kernel
\[K_N(x,y)=\ba
S_{N\ov2}(x-y)&DS_{N\ov2}(x-y)\\\\IS_{N\ov2}(x-y)-\ep(x-y)&S_{N\ov2}(x-y).
\ea\]\sp

\setcounter{equation}{0}\renewcommand{\theequation}{8.\arabic{equation}}
\begin{center}{\bf 8. The {\boldmath$\be=4$}\ Hermitian
ensembles}\end{center}\sp

We begin now with the $N$-fold integral
\[\int\cd\int\prod_{j<k}(x_j-x_k)^4\,\prod_jw(x_j)\,\prod_j(1+f(x_j))\,dx_1\cd
dx_N.\]
Using (\ref{pretty}) as it stands and (\ref{4det}) we see that
the square of this $N$-fold integral
equals a constant depending only on $N$ times
\[\det\,\Big(\int(k-j)\,x^{j+k-1}\,(1+f(x))\,w(x)\,dx\Big)_{j,k=0,\cd,2N-1}.\]

We replace the sequence $\{x^j\}$ by any sequence $\{p_j(x)\}$ of
polynomials of exact degree $j$. Except for another constant
factor depending only on $N$, the above equals
\[\det\,\Big(\int(p_j(x)\,p_k'(x)-p_j'(x)\,p_k(x))\,(1+f(x))\,w(x)\,dx\Big).\]
Just as when $\be=2$ it was convenient to introduce
functions $\ph_j$ which were equal to the orthonormal polynomials times
$w^{\hf}$
so now we introduce $\ps_j=p_j\,w^{\hf}$. The matrix is equal to
\[\Big(\int(\ps_j(x)\,\ps_k'(x)-\ps_j'(x)\,\ps_k(x))\,(1+f(x))\,dx\Big),\]
the extra terms arising from the differentiations having cancelled. We
write this as
\[M+\Big(\int(\ps_j(x)\,\ps_k'(x)-\ps_j'(x)\,\ps_k(x))\,f(x)\,dx\Big)\]
where $M$ is the matrix of integrals
$\int(\ps_j(x)\,\ps_k'(x)-\ps_j'(x)\,\ps_k(x))\,dx$.

Next we factor out $M$, say on the left. Its determinant is just another
constant
depending only on $N$. If
$M^{-1}=(\mu_{jk})$ and we set $\et_j(x)=\sum\,\mu_{jk}\,\ps_k(x)$ then
the resulting
matrix is
\[I+\Big(\int(\et_j(x)\,\ps_k'(x)-\et_j'(x)\,\ps_k(x))\,f(x)\,dx\Big),\]

Next, we write the sum of products
$\et_j(x)\,\ps_k'(x)-\et_j'(x)\,\ps_k(x)$ as a matrix product
\[\left(\begin{array}{cc}\et_j(x)&-\et_j'(x)\end{array}\right)
\,\left(\begin{array}{c}\ps_k'(x)\\\\\ps_k(x)\end{array}\right).\]
Thus if we set
\[A(j,x)=f(x)\,\left(\begin{array}{cc}\et_j(x)&-\et_j'(x)\end{array}\right),
\ \
B(x,j)=\left(\begin{array}{c}\ps_j'(x)\\\\\ps_j(x)\end{array}\right),\]
then the above matrix is $I+AB$. In this case $BA$ is the integral
operator with
matrix kernel $K_N(x,y)\,f(y)$ where
\[K_N(x,y)=\left(\begin{array}{cc}\sum \ps_j'(x)\,\et_j(y)&-\sum
\ps_j'(x)\,\et_j'(y)\\\\
\sum \ps_j(x)\,\et_j(y)&-\sum \ps_j(x)\,\et_j'(y)\end{array}\right).\]
The sums here are over $j=0,\cd,2N-1$.

We have shown that with this $K_N$ (\ref{1Pf}) holds up to a constant
factor depending
on $N$ and as usual taking $f=0$ shows that the constant factor equals
1.

Recall the definition of the $\et_j$. If we write
\[S_N(x,y)=2\sum_{j,k=0}^{2N-1} \ps_j'(x)\,\mu_{jk}\,\ps_k(y)\]
and
\[IS_N(x,y)=2\sum_{j,k=0}^{2N-1} \ps_j(x)\,\mu_{jk}\,\ps_k(y),
\ \ S_ND(x,y)=-2\sum_{j,k=0}^{2N-1} \ps_j'(x)\,\mu_{jk}\,\ps_k'(y)\]
then
\bq
K_N(x,y)=\hf\left(\begin{array}{cc}S_N(x,y)&S_ND(x,y)\\\\IS_N(x,y)&S_N(y,x)
\end{array}\right).\label{4K}\eq
The explanation for the notation is that if $S_N$ is the operator with
kernel $S_N(x,y)$
then $S_ND(x,y)$ is the kernel of $S_ND$ ($D=$ differentiation) and
$IS_N(x,y)$ is
the kernel of $IS_N$ ($I=$ integration, more or less). The factors 2 and
$\hf$ are
inserted to maintain the analogy with (\ref{4Kcirc}).

One might wonder about the choice of the $p_j$ and the resulting
$\mu_{jk}$. Since
$M$ is antisymmetric the formulas would be simplest if it were the
direct sum of $N$
copies of the $2\times2$ matrix $Z=\ba0&1\\-1&0\ea$,
whose inverse equals its negative. The polynomials that
achieve this are called skew-orthogonal.
But there is no actual necessity for using these since, as is easily
seen,
any family of polynomials leads to the same matrix kernel.\sp

\setcounter{equation}{0}\renewcommand{\theequation}{9.\arabic{equation}}
\begin{center}{\bf 9. The {\boldmath$\be=1$}\ Hermitian
ensembles}\end{center}\sp

Again we assume that $N$ is even and begin now with
\[\int\cd\int\prod_{j<k}|x_j-x_k|\,\prod_jw(x_j)\,\prod_j(1+f(x_j))\,dx_1\cd
dx_N.\]
This is equal to $N!$ times the integral over the region $x_1\leq\cd\leq
x_N$ and each
$|x_j-x_k|=x_k-x_j$ there. Using (\ref{1det}) we see therefore that the
square
of the integral equals a constant depending only on $N$ times
\[\det\Big(\int\int\ep(x-y)\,x^j\,y^k\,(1+f(x))\,
(1+f(y))\,w(x)\,w(y)\,dy\,dx\Big)
_{j,k=0,\cd,N-1}.\]
As in the last section we replace $\{x^j\}$ by an arbitrary sequence of
polynomials
$\{p_j(x)\}$ of exact degree $j$.. The above becomes a constant
depending
only on $N$ times
\[\det\Big(\int\int\ep(x-y)\,p_j(x)\,p_k(y)\,
(1+f(x)+f(y)+f(x)\,f(y))\,w(x)\,w(y)\,dy\,dx
\Big).\]

Now set $\ps_j(x)=p_j(x)\,w(x)$ and denote by $M$ the matrix with $j,k$
entry
\[\int\int\ep(x-y)\,p_j(x)\,p_k(y)\,w(x)\,w(y)\,dy\,dx.\]
Then the last determinant equals
\[\det\Big(M+\int\int\ep(x-y)\,\ps_j(x)\,\ps_k(y)\,
(1+f(x)+f(y)+f(x)\,f(y))\,dy\,dx
\Big)\]
\[=\det\Big(M+\int\,[f\,\ps_j\,
\ep\ps_k-f\,\ps_k\,\ep\ps_j-f\,\ps_k\,\ep(f\ps_j)]\,dx\Big).\]
All expressions in the integrand are functions of $x$. As earlier,
the operator with kernel $\ep(x-y)$ is denoted by $\ep$.

As in the last section, we factor out $M$ on the left, write
$M^{-1}=(\mu_{jk})$ and
set $\et_j=\sum\,\mu_{jk}\,\psi_k$. What results is the determinant of
\[I+\Big(\int\,[f\,\et_j\,\ep\ps_k
-f\,\ps_k\,\ep\et_j-f\,\ps_k\,\ep(f\et_j)]\,dx\Big),\]
and, as usual, at this point the multiplying constant equals 1.

The integral is equal to the $j,k$ entry of the product $AB$, where
\[A(j,x)=\left(\begin{array}{ccc}-f\,\ep\et_j-f\,
\ep(f\et_j)&&f\,\et_j\end{array}\right),
\ \ \
B(x,j)=\left(\begin{array}{c}\ps_j\\\\\ep\ps_j\end{array}\right).\]
The operator $I+BA$ equals
\[\ba I-\sum \ps_j\tn f\,\ep\et_j-\sum \ps_j\tn f\,\ep(f\et_j)&
\sum \ps_j\tn f\,\et_j\\&
\\-\sum \ep\ps_j\tn f\,\ep\et_j-\sum \ep\ps_j\tn f\,\ep(f\et_j)&
I+\sum \ep\ps_j\tn f\,\et_j\ea.\]
As in the circular ensemble, we can write this as the matrix product
\[\ba I-\sum \ps_j\tn f\,\ep\et_j&\sum \ps_j\tn f\,\et_j\\&
\\-\sum \ep\ps_j\tn f\,\ep\et_j-\ep f&I+\sum \ep\ps_j\tn f\,\et_j\ea
\ \ba I&0\\\\\ep f&I\ea.\]
The determinant of the product equals the determinant of the first
factor, i.e., the
determinant of
\[I+\ba -\sum \ps_j\tn \ep\et_j&\sum \ps_j\tn \et_j\\&
\\-\sum \ep\ps_j\tn \ep\et_j-\ep&\sum \ep\ps_j\tn \et_j\ea\,f.\]
Recalling the definition of the $q_j$ we now define
\[S_N(x,y)=-\sum_{j,k=0}^{N-1}\ps_j(x)\,\mu_{jk}\,\ep\ps_k(y),\]
\[IS_N(x,y)=-\sum_{j,k=0}^{N-1}\ep\ps_j(x)\,\mu_{jk}\,\ep\ps_k(y),
\ \ S_ND(x,y)=\sum_{j,k=0}^{N-1}\ps_j(x)\,\mu_{jk}\,\,\ps_k(y),\]
and we see that the (\ref{1Pf}) holds for this ensemble when $K_N$ has
kernel
\bq
K_N(x,y)=\left(\begin{array}{cc}S_N(x,y)&
S_ND(x,y)\\\\IS_N(x,y)-\ep(x-y)&S_N(y,x)
\end{array}\right).\label{1K}\eq

Once again the matrix kernel is independent
of the choice of the $p_j$ but the expression is simplest when the
polynomials
are skew-orthogonal in this context, i.e., when $M$ is the direct sum of
$N/2$
copies of $Z$.\sp

\setcounter{equation}{0}\renewcommand{\theequation}{10.\arabic{equation}}
\begin{center}{\bf 10. The Gaussian ensembles}\end{center}\sp

\noi{\it The Gaussian symplectic ensemble ($\be=4$)}. The weight
function $w(x)$ now equals
$e^{-x^2}$. As in sec.~5 we use the notation $\ph_j$
for the polynomials orthonormal with respect to the weight function
times the square root
of the weight function. It turns out that in this case the
skew-orthogonal polynomials
are simply expressed in terms of the $\ph_j$. In fact, in the notation
of sec.~8, we have
\[ \ps_{2n+1}=\ph_{2n+1}/\sqrt2,\ \ \ \ps_{2n}=-\ep\ph_{2n+1}/\sqrt2.\]
In other words,
\[p_{2n+1}(x)=e^{x^2/2}\,\ph_{2n+1}(x)/\sqrt2,\ \ \
p_{2n}(x)=-e^{x^2/2}\,
\ep\ph_{2n+1}/\sqrt2.\]
The point is that $p_{2n}$ as so defined is actually a polynomial of
degree $2n$, as is
easily seen. The skew-orthogonality is easy. If $j$ and $k$ are of the
same parity the
corresponding matrix entry equals 0, and if they have opposite parity we
compute, (using
the fact that $D\ep=$ Identity)
\[\int (p_{2n}(x)\,p_{2m+1}'(x)-p_{2n}'(x)\,p_{2m+1}(x))\,e^{-x^2}\,dx
=\int (\ps_{2n}\,\ps_{2m+1}'-\ps_{2n}'\,\ps_{2m+1})\,dx\]
\[=\hf\,\int (\ph_{2n+1}\,\ph_{2m+1}-\ep\ph_{2n+1}\,\ph_{2m+1}')\,dx.\]
Integration by parts applied to the second integrand and the
orthonormality of the $\ph_j$
show that the above equals $\dl_{n,m}$.

We now compute the entries of the matrix (\ref{4K}). If we keep in mind
that the inverse
of $M$ in this case equals $-M$ we see that $S_N(x,y)$ is equal to $1/2$
times
\bq\sum_{n=0}^{N-1}\ph_{2n+1}(x)\,\ph_{2n+1}(y)-\sum_{n=0}^{N-1}\ph_{2n+1}'(x)
\;\ep\ph_{2n+1}(y).\label{SN}\eq
To compute the second sum we use the differentiation formula
\[\ph_i'=\sqrt{{i\ov2}}\,\ph_{i-1}-\sqrt{{i+1\ov2}}\,\ph_{i+1}.\]
If $(a_{ij})$ is the antisymmetric tridiagonal matrix with $a_{i,\;i-1}=
\sqrt{i/2}$ then
\bq\ph_i'=\sum_{j\geq0}a_{ij}\,\ph_j,\ \ \
\ph_j=\sum_{i\geq0}a_{ji}\;\ep\ph_i.\label{phidiff}\eq
The first is just a restatement of the differentiation formula and the
second follows
from the first by applying $\ep$ and interchanging $i$ and $j$.
By the first part of (\ref{phidiff}) the second sum in (\ref{SN})
equals $\sum a_{ij}\,\ph_j\cdot\ep\ph_i$ summed over all $i,j\geq0$
such that $i$ is odd and $\leq 2N-1$. Using the antisymmetry of
$(a_{ij})$ and
a different description of the range of summation, we see that this
equals
$-\sum \ph_j\cdot a_{ji}\,\ep\ph_i$ summed over all $i,j\geq0$ such that
$j$ is even
and $\leq 2N$ except for the single term corresponding to $i=2N+1,\
j=2N$. (Recall that
$a_{ij}=0$ unless $|i-j|=1$.) By the second part of
(\ref{phidiff}) this equals
\bq-\sum_{n=0}^{N}\,\ph_{2n}(x)\,\ph_{2n}(y)-
\sqrt{N+\hff}\;\ph_{2N}(x)\;\ep\ph_{2N+1}(y).
\label{phisum}\eq

Recall that $S_N(x,y)$ equals $1/2$ times the sum (\ref{SN}). If we now
denote by
$S_N(x,y)$ the sum itself, then
\[S_N(x,y)=\sum_{n=0}^{2N}\,\ph_{n}(x)\,\ph_{n}(y)+\sqrt{N+\hff}\;\ph_{2N}(x)\;
\ep\ph_{2N+1}(y),\]
and consequently
\[IS_N(x,y)=\sum_{n=0}^{2N}\,\ep\ph_{n}(x)\;
\ph_{n}(y)+\sqrt{N+\hff}\;\ep\ph_{2N}(x)\;
\ep\ph_{2N+1}(y),\]
\[S_ND(x,y)=-\sum_{n=0}^{2N}\,\ph_{n}(x)\,
\ph_{n}'(y)-\sqrt{N+\hff}\;\ph_{2N}(x)\,
\ph_{2N+1}(y).\]
(That $IS_N(x,y)$ is obtained by applying $\,\ep$ to
(\ref{SN}) as a function of $x$ is a consequence of the fact that only
odd indices
occur there.) Hence our matrix kernel is given by
\[
K_N(x,y)=\hf\,\left(\begin{array}{cc}S_N(x,y)&S_ND(x,y)\\\\IS_N(x,y)&S_N(y,x)
\end{array}\right),\]
in analogy with (\ref{4Kcirc}).\sp

\noi{\it The Gaussian orthogonal ensemble ($\be=1$)}. In order to
continue using the same
functions $\ph_j$ as before we shall now take
as our weight function $w(x)=e^{-x^2/2}$. Thus the functions $\ps_j$ of
sec.~9 will
be of the same form, polynomial times $e^{-x^2/2}$. For the $p_j$ to be
skew-orthogonal
in the present context the antisymmetric matrix $M$ with $j,k$ entry
\[\int\int \ep(x-y)\,\ps_j(x)\,\ps_k(y)\,dy\,dx=\int
\ps_j(x)\;\ep\ps_k(x)\,dx\]
must be a direct sum of $N/2$ copies of $Z$. (Recall that $N$ is even.)
The $\ps_j$
are again very simple. They are given by
\[ \ps_{2n}=\ph_{2n},\ \ \ \ps_{2n+1}=\ph_{2n}'.\]
That these give polynomials of the right degree (and parity) is clear,
and we compute
\[\int \ps_{2n}(x)\;\ep\ps_{2m+1}(x)\,dx=\int
\ph_{2n}(x)\,\ph_{2m}(x)\,dx=\dl_{n,m}.\]

The sum $S_N(x,y)$ in (\ref{1K}) is now given by
\[\sum_{n=0}^{{N\ov2}-1}(\ps_{2n}(x)\;\ep\ps_{2n+1}(y)-
\ps_{2n+1}(x)\;\ep\ps_{2n}(y))
=\sum_{n=0}^{{N\ov2}-1}\ph_{2n}(x)\,\ph_{2n}(y)-
\sum_{n=0}^{{N\ov2}-1}\ph_{2n}'(x)
\;\ep\ph_{2n}(y).\]
By (\ref{phidiff}) the second sum on the right, equals $\sum
a_{ij}\,\ph_j\cdot\ep\ph_i$
summed over all $i,j\geq0$ such that $i$ is even and $\leq N-2$. This
equals
$-\sum \ph_j\cdot a_{ji}\,\ep\ph_i$ summed over all $i,j\geq0$ such that
$j$ is odd
and $\leq N-1$ except for the single term corresponding to $i=N,\
j=N-1$. This sum equals
\[-\sum_{n=0}^{{N\ov2}}\,\ph_{2n-1}(x)\,\ph_{2n-1}(y)-
\sqrt{{N\ov2}}\;\ph_{N-1}(x)\;
\ep\ph_{N}(y).\]
Hence we have now
\[S_N(x,y)=\sum_{n=0}^{N-1}\,\ph_{n}(x)\,
\ph_{n}(y)+\sqrt{{N\ov2}}\;\ph_{N-1}(x)\;
\ep\ph_{N}(y),\]
and consequently the other entries of the matrix $K_N(x,y)$ in
(\ref{1K}) are given by
\[IS_N(x,y)=\sum_{n=0}^{N-1}\,\ep\ph_{n}(x)\;
\ph_{n}(y)+\sqrt{{N\ov2}}\;\ep\ph_{N-1}(x)\;
\ep\ph_{N}(y),\]
\[S_ND(x,y)=-\sum_{n=0}^{N-1}\,\ph_{n}(x)\,
\ph_{n}'(y)-\sqrt{{N\ov2}}\;\ph_{N-1}(x)\,
\ph_{N}(y).\]

\begin{center}{\bf Acknowledgements}\end{center}

The work of the first author was supported in part by the National
Science Foundation through Grant DMS--9303413, and the work of the
second
author was supported in part by the National Science Foundation through
Grant DMS--9424292.\sp

\end{document}